%%%%%%%%%%%%%%%%%%%%%%%%%%%%%%%%%%%%%%%%%%%%%%%%%%%%%%%%%%%%%%%%%%%%%%%%%%%%
%%%%                                                                    %%%%
%%%% CHERN-SIMONS INVARIANT FOR SO(1,3)-SEN CONNECTIONS                 %%%%
%%%%%%%%%%%%%%%%%%%%%%%%%%%%%%%%%%%%%%%%%%%%%%%%%%%%%%%%%%%%%%%%%%%%%%%%%%%%

\baselineskip 14pt plus 2pt

\font\lbf=cmbx10 scaled\magstep1
\def\sq{\hfill \vbox {\hrule width3.2mm \hbox {\vrule height3mm \hskip 2.9mm
\vrule height3mm} \hrule width3.2mm }} 
\def\ni{\noindent}

\def\ua{\underline a \,}          \def\ba{\bf a \,} 
\def\ub{\underline b \,}          \def\bb{\bf b \,} 
\def\uc{\underline c \,}          \def\bc{\bf c \,} 
\def\ud{\underline d \,}          \def\bd{\bf d \,}

          \def\bi{\bf i \,} 
          \def\bj{\bf j \,} 
\def\uk{\underline k \,}          \def\bk{\bf k \,} 
           
\def\um{\underline m \,}           
\def\un{\underline n \,}

%%%%%%%%%%%%%%%%%%%%%%%%%%%%%%%%%%%%%%%%%%%%%%%%%%%%%%%%%%%%%%%%%%%%%%%%%%%
%%%%%%%%%%%%%%%%%%%%%%%%%%%%%%%%%%%%%%%%%%%%%%%%%%%%%%%%%%%%%%%%%%%%%%%%%%%
\centerline{\lbf On a global conformal invariant of initial data sets}
\bigskip
\centerline{Robert Beig}
\centerline{Institut f\"ur Theoretische Physik}
\centerline{Universit\"at Wien}
\centerline{Boltzmanngasse 5, A-1090 Wien, \"Osterreich}
\centerline{e-mail: beig@pap.univie.ac.at}
\medskip
\centerline{L\'aszl\'o B Szabados}
\centerline{Research Institute for Particle and Nuclear Physics}
\centerline{H--1525 Budapest 114, P.O.Box 49, Hungary}
\centerline{e-mail: lbszab@rmki.kfki.hu}

\bigskip
\bigskip
\ni
In the present paper a global conformal invariant $Y$ of a closed initial 
data set is constructed. A spacelike hypersurface $\Sigma$ in a Lorentzian 
spacetime naturally inherits from the spacetime metric a differentiation 
${\cal D}_e$, the so-called real Sen connection, which turns out to be 
determined completely by the initial data $h_{ab}$ and $\chi_{ab}$ induced 
on $\Sigma$, and coincides, in the case of vanishing second fundamental 
form $\chi_{ab}$, with the Levi-Civita covariant derivation $D_e$ of the 
induced metric $h_{ab}$. $Y$ is built from the real Sen connection ${\cal 
D}_e$ in the similar way as the standard Chern--Simons invariant is built 
from $D_e$. The number $Y$ is invariant with respect to changes of $h_{ab}$ 
and $\chi_{ab}$ corresponding to conformal rescalings of the {\it spacetime} 
metric. In contrast the quantity $Y$ built from the {\it complex} Ashtekar 
connection is {\it not} invariant in this sense. The critical points of our 
$Y$ are precisely the initial data sets which are locally imbeddable into 
conformal Minkowski space. \par

\bigskip
\bigskip
\ni
{\lbf 1.  Introduction}\par

\bigskip
\bigskip
\ni
In general relativity 3-manifolds play a distinguished role since in the 
initial value formulation of the Einstein theory the initial data, a 
metric $h_{ab}$ and a symmetric tensor field $\chi_{ab}$, are defined on 
connected orientable 3-manifolds $\Sigma$.  Because of the complexity of 
the initial value formulation, any invariant characterization of the 
initial data could provide a deeper understanding of the dynamics of general 
relativity. Mathematicians have extensively studied the geometry and 
the invariant characterization of three dimensional Riemannian manifolds. In 
particular, Chern and Simons [1] introduced a global conformal invariant of 
closed, orientable Riemannian 3-manifolds as the integral of the so-called 
Chern--Simons 3-form built from the Levi-Civita connection. The stationary 
points of this integral, viewed as a functional of the 3-metric, are 
precisely the conformally flat 3-geometries. Thus it might be interesting 
to generalize this result for initial data sets of general relativity, 
obtaining a global conformal invariant of the initial data; and it might 
also be interesting even from a pure mathematical point of view if there is 
a similar conformal invariant for connections on other trivial principal 
or vector bundles over $\Sigma$. \par
        In the present paper we show that the Chern--Simons functional, 
built from the real Sen connection on a four dimensional trivializable 
Lorentzian vector bundle over a closed orientable 3-manifold $\Sigma$, is 
invariant with respect to rescalings of $h_{ab}$ and $\chi_{ab}$ 
corresponding to {\it spacetime} conformal rescalings; and the stationary 
points of this functional are precisely those triples $(\Sigma,h_{ab},
\chi_{ab})$ that can be locally imbedded into some conformally flat 
Lorentzian spacetime with first and second fundamental forms $h_{ab}$ and 
$\chi_{ab}$, respectively. For time symmetric initial data, i.e. when 
$\chi_{ab}=0$, our invariant reduces to that of Chern and Simons, i.e. our
invariant is a natural generalization of the latter. \par
       The second section of this paper is a review of the most important 
properties of the Chern--Simons functional for a general gauge group and the 
specific conformal invariant of Chern and Simons for Riemannian 3-manifolds.  
The way in which we introduce their invariant, however, is slightly 
different from the original one, because it is this way that can be
generalized to find Chern--Simons invariants for gauge groups larger than 
the rotation group. \par
         The third section is devoted to the Chern--Simons invariant of a 
triple $(\Sigma,h_{ab},\chi_{ab})$. We denote our invariant by $Y$. 
Although in this paper we will not use the Einstein equations (or any other 
field equations), for the sake of simplicity we call such a triple an 
initial data set. First we consider a trivializable Lorentzian vector bundle 
$V(\Sigma)$ over $\Sigma$, introduce the real Sen connection on it and then 
the Chern--Simons functional, built from the real Sen connection, will 
be introduced. In the third subsection we clarify some of its properties, 
in particular its conformal invariance, and we calculate its variational 
derivatives. For the sake of completeness in the fourth subsection we 
consider the Chern--Simons functional built from the {\it complex} 
Ashtekar connection. We consider this connection as a connection on the 
bundle of self-dual 2-forms on the Lorentzian vector bundle determined 
by the real Sen connection. It turns out, however, that this 
Ashtekar--Chern--Simons functional is {\it not} invariant with respect to 
conformal rescalings. Thus the conformal invariance depends on the 
representation in which the Chern--Simons functional is constructed. In 
fact, the stationary points of this functional are the initial data sets 
that can be locally isometrically imbedded into a flat spacetime. \par
          The fourth section is devoted to the local isometric 
imbeddability of initial data sets into conformally flat geometries. More 
precisely, if $\Sigma$ is an $n$ dimensional manifold ($n\geq3$), $h_{ab}$ 
a metric on $\Sigma$ with signature $(p,q)$, $p+q=n$, and $\chi_{ab}$ is a 
symmetric tensor field on $\Sigma$, then we are interested in the necessary 
and sufficient conditions for the triple $(\Sigma,h_{ab},\chi_{ab})$ to be 
locally isometrically imbeddable into some conformally flat $(n+1)$ 
dimensional geometry $(M,g_{ab})$ with $g_{ab}$ of signature $(p+1,q)$ or 
$(p,q+1)$ and so that the induced metric and second fundamental form are 
$h_{ab}$ and $\chi_{ab}$, respectively. We find three tensor fields, built 
from $h_{ab}$ and $\chi_{ab}$, whose vanishing characterizes this local 
imbeddability. In three dimensions one of these tensor fields vanishes 
identically, and the remaining two are given by the variational derivatives 
of the Sen--Chern--Simons functional. Thus the stationary points of the 
Sen--Chern--Simons functional are precisely the initial data sets that can 
be imbedded, at least locally, into some conformally flat spacetime.  \par
       Ultimately, one wants to use the results of this paper in the study 
of solutions to the Einsten equations. An obvious question is that of the 
dependence of $Y$ on $(h_{ab},\chi_{ab})$, when the latter is evolved via 
the Einstein vacuum equations. This topic will be addressed in work in 
progress. \par 
       Our conventions are mostly the same as those of [2]. In particular, 
the wedge product of forms is defined to be the anti-symmetric part of the 
tensor product, the signature of the spacetime and spatial metrics 
is (+ -- -- --) and (-- -- --), respectively. The curvature $F^a{}_{b\alpha
\beta}$ of a covariant derivation $\nabla_\alpha$ on a vector bundle is 
defined by $-F^a{}_{b\alpha\beta}X^bv^\alpha w^\beta:=v^\alpha\nabla_\alpha
(w^\beta\nabla_\beta X^a)-w^\beta\nabla_\beta(v^\alpha\nabla_\alpha X^a)-
[v,w]^\alpha\nabla_\alpha X^a$. Finally, the Ricci tensor  is $R_{ab}:=R^e
{}_{aeb}$ and the curvature scalar is the contraction of $R_{ab}$ with the 
metric. Although we mostly use the abstract index notation, sometimes the 
differential form notation will also be used. Every mapping, section, tensor 
field, etc. will be smooth. Our general differential geometric reference is 
[3].

\bigskip 
\bigskip
\ni 
{\lbf 2 The Chern--Simons functional}\par 
\bigskip 
\ni 
{\bf 2.1 The general Chern--Simons functional}\par 
\bigskip 
\ni 
Let $G$ be any Lie group, ${\cal G}$ its Lie algebra and $\pi:P\rightarrow 
\Sigma$ a trivializable principal fibre bundle over $\Sigma$ with structure 
group $G$. Since $P$ is trivializable, it admits global cross sections 
$\sigma:\Sigma\rightarrow P$. Let $V$ be a $k$ dimensional real vector space, 
$\rho:G\rightarrow{\rm GL}(V)$ a linear representation of $G$ on $V$, 
$\rho_\ast:{\cal G}\rightarrow{\rm gl} (V)$ the corresponding representation 
of its Lie algebra and let $E(\Sigma)$ be the vector bundle over $\Sigma$ 
associated to $P$ with the linear representation $\rho$ of $G$ on $V$. 
Because of the trivializability of $P$, $E(\Sigma)$ is also trivializable, 
and hence it admits $k$ pointwise linearly independent global sections $e^a
_{\ua}$, ${\ua}=1, ...,k$.  We call such a system of global sections a 
global frame field.  Any global cross section of $P$ can be interpreted as 
such a global frame field, and the `gauge transformations' as certain 
$k\times k$-matrix valued functions $\Lambda^{\ua}{}_{\ub}$ on $\Sigma$.
\par 
        Any connection on $P$ determines a connection on $E(\Sigma)$, whose 
connection coefficients with respect to a global frame field form a $\rho 
_\ast({\cal G})\subset{\rm gl}(k,{\bf R})$-valued 1-form $A^{\ua} _{\mu
{\ub}}$ on $\Sigma$. Here $\mu$ is the abstract tensor index referring to 
the manifold $\Sigma$. If $-F^{\ua}{}_{{\ub}\mu\nu}:=\partial_\mu A^{\ua}
_{\nu{\ub}}-\partial_\nu A^{\ua}_{\mu{\ub}}+A^{\ua}_{\mu{\uc}}A^{\uc}_{\nu
{\ub}}-A^{\ua}_{\nu{\uc}}A^{\uc}_{\mu{\ub}}$, the curvature of the 
connection on $\Sigma$, then the Chern--Simons functional of the connection 
is the integral

$$ 
Y[A]:=\int_\Sigma {\rm Tr}\Bigl(F_{[\mu\nu}A_{\rho]}+{2\over3}A_{[\mu}
A_\nu A_{\rho]}\Bigr):=
\int_\Sigma\Bigl(F^{\ua}{}_{{\ub}\alpha\beta}A^{\ub}_{\gamma{\ua}}
+{2\over3}A^{\ua}_{\alpha{\ub}}A^{\ub}_{\beta{\uc}}A^{\uc}_{\gamma{\ua}}
\Bigr){1\over3!}\delta^{\alpha\beta\gamma}_{\mu\nu\rho}.\eqno(2.1.1)
$$
\ni
Obviously, $Y[A]$ is invariant with respect to orientation preserving 
diffeomorphisms of $\Sigma$ onto itself. Recalling that under a gauge 
transformation $\Lambda^{\ua}{}_{\ub}$ the connection and the curvature 
transform as

$$\eqalign{
A^{\ua}_{\mu{\ub}}\mapsto A^\prime{}^{\ua}_{\mu{\ub}}&:=\Lambda_{\ud}{}^{\ua}
 \Bigl(A^{\ud}_{\mu{\uc}}\Lambda^{\uc}{}_{\ub}+\partial_\mu\Lambda^{\ud}
 {}_{\ub}\Bigr),\cr 
F^{\ua}{}_{{\ub}\mu\nu}\mapsto F^{\prime{\ua}}{}_{{\ub}\mu\nu}&:=
 \Lambda_{\ud}{}^{\ua}F^{\ud}{}_{{\uc}\mu\nu}\Lambda^{\uc}{}_{\ub},\cr} 
$$
\ni
where $\Lambda_{\ub}{}^{\uc}$ is defined by $\Lambda^{\ua}{}_{\uc}\Lambda
_{\ub}{}^{\uc}=\delta^{\ua}_{\ub}$, the Chern--Simons functional transforms 
as

$$\eqalign{ 
Y[A]-Y[A^\prime]&=2\int_\Sigma\partial_\alpha\Bigl(A^{\ua}_{\beta{\ub}}
 \Lambda_{\ua}{}^{\uc}\bigl(\partial_\gamma\Lambda^{\ub}{}_{\uc}\bigr)
 \Bigr){1\over3!}\delta^{\alpha\beta\gamma}_{\mu\nu\rho}+\cr
&+{2\over3}\int_\Sigma \Lambda_{\uk}{}^{\ua}\bigl(\partial_\alpha\Lambda
 ^{\um}{}_{\ua}\bigr)\Lambda_{\um}{}^{\ub}\bigl(\partial_\beta\Lambda
 ^{\un}{}_{\ub}\bigr)\Lambda_{\un}{}^{\uc}\bigl(\partial_\gamma\Lambda
 ^{\uk}{}_{\uc}\bigr){1\over3!}\delta^{\alpha\beta\gamma}_{\mu\nu\rho}.\cr}
\eqno(2.1.2)
$$
\ni
Its first term on the right is zero by $\partial\Sigma=\emptyset$.  First 
suppose that the gauge transformation $\Lambda^{\ua}{}_{\ub}$ is homotopic 
to the identity transformation, i.e. there is a 1 parameter family of global 
gauge transformations $\Lambda^{\ua}{}_{\ub}(t)$, $t\in [0,1]$, such that
$\Lambda^{\ua}{}_{\ub}(0)=\delta^{\ua}_{\ub}$ and $\Lambda^{\ua}{}_{\ub}(1)
=\Lambda^{\ua}{}_{\ub}$ (`small gauge transformations'). Then substituting 
$\Lambda^{\ua}{}_{\ub}(t)$ into (2.1.2) and taking the derivative with 
respect to $t$ at $t=0$ we obtain that the right hand side is vanishing; 
i.e. the Chern--Simons functional is invariant with respect to small gauge 
transformations. For general gauge transformations, however, the second 
term on the right of (2.1.2) is not zero. In fact, as a consequence of the 
integrality of the second Chern class, for the left hand side of (2.1.2) we 
have (see e.g. [4])

$$
Y[A]-Y[A^\prime]=16\pi^2 N, \eqno(2.1.3) 
$$
\ni
for some integer $N$ depending on the global gauge transformation $\Lambda
^{\ua}{}_{\ub}$. We will see that the geometric content of this formal 
result is connected with a certain homotopy invariant of the mapping 
$\Lambda:\Sigma\rightarrow G$. In particular, for $G=SO(3)$ or $SO_0(1,3)$, 
the connected component of $SO(1,3)$, $N$ is just the integer that can be 
interpreted as twice the winding number of $\Lambda$. \par
      Finally, let us consider any smooth 1 parameter family $A^{\ua}_{\mu
{\ub}}(t)$ of connections on $E(\Sigma)$ and the corresponding Chern--Simons 
functional $Y[A(t)]$. Then

$$
\delta Y[A]:=\Bigl({{\rm d}\over{\rm d}t}Y[A(t)]\Bigr)\vert_{t=0}=2\int
_\Sigma\Bigl({\rm Tr}\bigl(F_{[\mu\nu}\delta A_{\rho]}\bigr)+\partial_{[\mu}
\bigl({\rm Tr}\bigl(A_\nu\delta A_{\rho]}\bigr)\bigr)\Bigr), \eqno(2.1.4) 
$$
\ni
where $\delta A^{\ua}_{\mu{\ub}}:=({{\rm d}\over{\rm d}t}A^{\ua}_{\mu{\ub}}
(t))\vert_{t=0}$, the `variation' of the connection 1-form. Thus $Y[A]$ is 
functionally differentiable and the derivative is essentially the curvature. 
\bigskip
\bigskip

\ni
{\bf 2.2 The conformal invariant of Chern and Simons for Riemannian
         3-manifolds}\par
\bigskip
\ni
Let $R\rightarrow\Sigma$ be a trivializable principal bundle over $\Sigma$ 
with structure group $SO(3)$, $\rho$ the defining representation of $SO(3)$ 
and let $E(\Sigma)$ be the associated trivializable vector bundle. The global 
sections of $R$ can be interpreted as globally defined frame fields $E^a
_{\bi}$ of $E(\Sigma)$, ${\bi}=1,2,3$. Then one can introduce the negative 
definite fibre metric $h_{ab}$ for which $E^a_{\bi}$ (and hence any frame 
field obtained from $E^a_{\bi}$ by the action of $SO(3)$) is orthonormal;
i.e.  if $\vartheta ^{\bi}_a$ is the basis dual to $E^a_{\bi}$ and $\eta
_{\bi\bj}:={\rm diag} (-1,-1,-1)$, then $h_{ab}:=\vartheta^{\bi}_a
\vartheta^{\bj}_b\eta_{\bi\bj}$. (The dual basis can also be interpreted as 
a vector bundle isomorphism $E(\Sigma)\rightarrow\Sigma\times{\bf R}^3:$ 
$(p,X^a)\mapsto(p,X^{\bi})$ and the fibre metric $h_{ab}$ is the pull back 
of the constant metric $\eta_{\bi\bj}$ along $\vartheta^{\bi}_a$.) Any 
connection on $R$ determines a covariant derivation $D_\alpha$ on 
$E(\Sigma)$, annihilating the fibre metric $h_{ab}$. If $E^a_{\bi}$, 
$\vartheta^{\bi}_a$ is a pair of dual global $h_{ab}$-orthonormal frame 
fields, then the connection can be characterized completely by its 
connection coefficients $\gamma^{\bi}_{\alpha{\bj}}:=\vartheta^{\bi}_aD
_\alpha E^a_{\bj}$.\par 
      Since for any orientable 3-manifold the tangent bundle is trivializable, 
there is a vector bundle isomorphism, the so-called soldering form, 
between the tangent bundle and the abstract vector bundle $E(\Sigma)$. 
It is $\theta: T\Sigma\rightarrow E(\Sigma):$ $(p,v^\alpha)\mapsto(p,
v^\alpha\theta^a_\alpha)$. By means of the soldering form $T\Sigma$ and 
$E(\Sigma)$ can be identified (and there will not be any difference between 
the Greek and Latin indices) and $h_{ab}$ will be a metric on $T\Sigma$. 
Obviously, for any fixed soldering form, there is a one-to-one 
correspondence between the negative definite metrics on $T\Sigma$ and the 
global frame fields in $E(\Sigma)$ modulo the $SO(3)$ action. Furthermore,
the connection on $E(\Sigma)$ determines a linear metric connection on
$T\Sigma$. Requiring the vanishing of the torsion of this linear connection,
the connection on $E(\Sigma)$ will be completely determined and the
connection coefficients $\gamma^{\bi}_{\alpha{\bj}}$ become the Ricci 
rotation coefficients of the Levi-Civita connection. Thus the Chern--Simons
functional, built from that connection on $E(\Sigma)$ whose pull back to
$T\Sigma$ is the Levi-Civita one, is completely determined by $E^a _{\bi}$.
Consequently, for such connections $Y$ will be a {\it second} order 
functional of $E^a_{\bi}$, invariant with respect to homotopically trivial 
gauge transformations, but it will depend on the homotopy class of the global 
frame field on $\Sigma$. Therefore $h_{ab}$ determines $Y[E^a_{\bi}]$ modulo 
$16\pi^2$ only. \par 
      To understand the root of this obstruction, recall that $\Lambda: 
\Sigma\rightarrow SO(3)$ is a proper map (i.e.  the inverse image of any 
compact subset of $SO(3)$ is compact, because $\Sigma$ itself is compact) 
and $\dim\Sigma=\dim SO(3)$. Thus there is an {\it integer}, ${\rm deg}
(\Lambda)$, the degree of $\Lambda$, such that for any 3-form $\omega$ on 
$SO(3)$ $\int_\Sigma\Lambda^\ast(\omega)={\rm deg}(\Lambda)\int_{SO(3)}
\omega$ [5]. In particular, for the normalized invariant volume element of 
$SO(3)$, ${\rm d}v:={1\over48\pi^2}{\rm Tr}((\lambda^{-1}{\rm d}\lambda)
\wedge(\lambda^{-1}{\rm d}\lambda)\wedge(\lambda^{-1}{\rm d}\lambda))$, by 
(2.1.2) and (2.1.3) we have 

$$
{\rm deg}(\Lambda)=\int_\Sigma\Lambda^\ast({\rm d}v)={1\over48\pi^2}\int
_\Sigma{\rm Tr}\Bigl(\bigl(\Lambda^{-1}{\rm d}\Lambda\bigr)\wedge\bigl(
\Lambda^{-1}{\rm d}\Lambda\bigr)\wedge\bigl(\Lambda^{-1}{\rm d}\Lambda\bigr)
\Bigr)={1\over2}N. \eqno(2.2.1)
$$
\ni
But ${\rm deg}(\Lambda)$ counts how many times $\Sigma$ covers the rotation 
group by the mapping $\Lambda$, and hence $N$ may be interpreted as twice 
the winding number of the map $\Lambda:\Sigma\rightarrow SO(3)$. In 
particular, for $\Sigma\simeq S^3$ the homotopy classes of the mapping 
$\Lambda$ are precisely the elements of $\pi_3(SO(3))$.\par 
       A direct calculation shows that $Y[E^a_{\bi}]$ is invariant with 
respect to the conformal rescalings $E^a_{\bi}\mapsto\Omega^{-1}E^a_{\bi}$, 
and hence $Y[E^a_{\bi}]$ modulo $16\pi^2$ is a conformal invariant of
$(\Sigma,h_{ab})$. Any 1-parameter family $E^a_{\bi}(t)$ of global frame
fields yields a 1-parameter family $\gamma^{\bi}_{a{\bj}}(t)$ of connection
coefficients, i.e. any variation $\delta E^a_{\bi}$ determines a variation
$\delta\gamma^{\bi}_{a{\bj}}$. Thus any variation of the metric $h_{ab}$
determines the variation of the connection coefficients, apart form an
unspecified {\it small} gauge transformation. Then by (2.1.4) it is a
straightforward calculation to show that the variational derivative of
$Y[E^a_{\bi}]$ with respect to $h_{ab}$ is well defined and it is the 
Cotton--York tensor [1]. Thus the stationary points of the $SO(3)$ 
Chern--Simons functional are, in fact, the conformally flat Riemannian 
metrics. It is this picture that we generalize in finding our conformal 
invariant of initial data sets in the next section.
\bigskip
\bigskip

\ni
{\lbf 3 The Chern--Simons invariant of initial data sets}\par
\bigskip
\ni
{\bf 3.1 The Lorentzian vector bundle}\par
\bigskip
\ni
Let $L\rightarrow\Sigma$ be a trivializable principal bundle over $\Sigma$ 
with the structure group $SO_0(1,3)$, $\rho$ its defining representation and 
let $V(\Sigma)$ be the associated vector bundle. $V(\Sigma)$ is therefore 
a trivializable real vector bundle of rank 4 over $\Sigma$. The global 
sections of $L$ can be considered as globally defined frame fields $e^a
_{\ua}$, ${\ua}=0,...,3$, with given `space' and `time' orientation; and one 
can define the Lorentzian fibre metric $g_{ab}$ on $V(\Sigma)$ for which 
$e^a_{\ua}$ is orthonormal. Explicitly, if $\zeta^{\ua}_a$ is the basis 
dual to $e^a_{\ua}$ and $\eta_{\ua\ub}:={\rm diag}(1,-1,-1,-1)$ then 
$g_{ab}:=\zeta^{\ua}_a\zeta ^{\ub}_b\eta_{\ua\ub}$. $\zeta^{\ua}_a$ can 
also be interpreted as a vector bundle isomorphism $V(\Sigma)\rightarrow
\Sigma\times{\bf R}^4:$ $(p,X^a)\mapsto(p,X^{\ua})$ and $g_{ab}$ as the 
pull back of $\eta_{\ua\ub}$ from $\Sigma\times{\bf R}^4$ to $V(\Sigma)$.
\par 
     Since both $T\Sigma$ and $V(\Sigma)$ are trivializable, there are 
imbeddings $\Theta:T\Sigma\rightarrow V(\Sigma):$ $(p,v^\alpha)\mapsto(p,
v^\alpha\Theta^a_\alpha)$ such that the vectors $v^\alpha\Theta^a_\alpha$ 
are all spacelike with respect to the fibre metric $g_{ab}$. Or, in other 
words, the pull back of $g_{ab}$ along $\Theta$, $h_{\alpha\beta}:=\Theta^a
_\alpha\Theta^b_\beta g_{ab}$, is a negative definite metric on $T\Sigma$. 
Thus $\Theta(T_p \Sigma)$ is a spacelike subspace of the fibre $V_p$ in 
$V(\Sigma)$ over $p\in\Sigma$, and hence, apart from a sign, there is a 
uniquely determined global section $t^a$ of $V(\Sigma)$ which has unit norm 
with respect to $g_{ab}$ and is a normal of $\Theta(T\Sigma)$: $v^\alpha
\Theta^a_\alpha t_a=0$ for all $v^\alpha$ tangent vector of $\Sigma$.
The orientation of $t^a$ will be chosen to be compatible with the `time'
orientation above. Then $P^a_b:=\delta^a_b-t^at_b$ is the projection of the
fibre $V_p$ onto $\Theta(T_p\Sigma)$ at each point $p$ of $\Sigma$. Thus if
$X^a$ is any section of $V(\Sigma)$ then it can be decomposed in a unique 
way as $X^a= Nt^a+N^a$, where $N$ is a function and $N^a$ is a section of 
$V(\Sigma)$ such that $P^a_bN^b=N^a$. $N$ and $N^a$ may be called the lapse 
and shift parts of $X^a$, respectively. Obviously, this decomposition 
depends on the imbedding $\Theta$. Any such decomposition of the sections 
of $V(\Sigma)$ into its lapse and shift parts defines a vector bundle 
isomorphism $\iota$ between $V(\Sigma)$ and the Whitney sum of the trivial 
line bundle $\Sigma\times{\bf R}$ and $T\Sigma$. For fixed $\Theta$ we can, 
and in fact we will, identify the tangent bundle $T\Sigma$ with its 
$\Theta$-image in $V(\Sigma)$. Then the Greek indices become 
$P^a_b$-projected Latin indices. In spite of this identification we use the 
Greek indices if we want to emphasize that they are indices tangential to 
$\Sigma$. Obviously, the negative definite metric $h_{ab}$ does not fix the 
Lorentzian fibre metric $g_{ab}$ completely: $g_{ab}$ and $\tilde g_{ab}$ 
determine the same spatial metric iff $\tilde g_{ab}=g_{ab}+\tau t_at_b$, 
where $\tau:\Sigma\rightarrow(-1,\infty)$ is an arbitrary function. If $X^a$ 
is any section of $V(\Sigma)$ then, under the transformation $g_{ab}\mapsto 
g_{ab}+\tau t_at_b$, its lapse part transforms as $N\mapsto\sqrt{1+\tau}N$, 
and hence this freedom corresponds to the pure rescaling of the lapse and 
the changing of the vector bundle isomorphism $\iota$ above. Thus the 
Lorentzian vector bundle $V(\Sigma)$ is completely determined by $h_{ab}$ 
and the knowledge of the lapse and shift parts of its sections. The vector 
bundle $V(\Sigma)$ can be interpreted as the restriction of the spacetime 
tangent bundle $TM$ to an imbedded spacelike hypersurface $\Sigma$, and 
$\Theta$ as the differential of the injection $\Sigma\rightarrow M$.\par
        A $g_{ab}$-orthonormal global frame field will be said to be 
compatible with the imbedding $\Theta$ if it is of the form $\{t^a,E^a_{\bi}
\}$, ${\bi}=1,2,3$. Thus $E^a_{\bi}$ is a triad of orthonormal vectors 
tangent to the distribution $\Theta(T\Sigma)$ everywhere. The set of all 
such $\Theta$-compatible frame fields defines a reduction $SO_0(1,3)
\rightarrow SO(3)$ of the gauge group (`time gauge'). As the next lemma 
shows, there is no topological obstruction excluding the possibility of 
such a gauge reduction.\par 
\medskip
\ni
{\bf Lemma 3.1.1:} For any global frame field $e^a_{\ua}$ there exists a 
globally defined one parameter family of Lorentz transformations $\Lambda 
(t):\Sigma\rightarrow SO_0(1,3)$, $t\in[0,1]$, such that $\Lambda^{\ua}
{}_{\ub}(0)=\delta^{\ua}_{\ub}$ and $\Lambda^{\ua}{}_{\ub}(1)$ takes $e^a
_{\ua}$ into a $\Theta$-compatible frame field. \par
\medskip
\ni
{\it Proof}: Because of the trivializability of $L$, there are globally 
defined Lorentz transformations taking $e^a_{\ua}$ into a $\Theta$-compatible 
global frame. These transformations are unique only up to spatial rotations 
keeping the normal $t^a$ fixed. Or, in other words, we search for global 
Lorentz transformations modulo rotations, i.e. an element of the coset 
space $SO_0(1,3)/SO(3)$ being homotopic to the identity. But $SO_0(1,3)/
SO(3)$ is homeomorphic to ${\bf R}^3$, which is a contractible topological 
space. Hence any two mappings $\Sigma\rightarrow SO_0(1,3)/ SO(3)$ are 
homotopic. In particular, there is a Lorentz transformation, taking $e^a
_{\ua}$ into a $\Theta$-compatible frame, which is homotopic to the 
identity transformation. \sq \par
\medskip
\ni
This Lemma implies that there is a natural one-to-one correspondence between 
the homotopy classes of the global rotations $\Sigma\rightarrow SO(3)$ and 
of the Lorentz transformations $\Sigma\rightarrow SO_0(1,3)$. \par
\bigskip
\bigskip

\ni
{\bf 3.2 The real Sen connection}\par
\bigskip
\ni
Any connection on $L$ determines a covariant derivation ${\cal D}_a$ on 
$V(\Sigma)$ which annihilates the Lorentzian fibre metric $g_{ab}$. However, 
we would like to build up our connection from the tensor fields $h_{ab}$, 
$\chi_{ab}$ of the initial data set. Thus we follow the philosophy of 
subsection 2.2 in tying the connection with the fields on $\Sigma$, and we 
specify ${\cal D}_a$ by imposing the following restrictions on its action 
on independent sections of $V(\Sigma)$. 
\item{i.} For the normal section $t_a$ let us define $\chi_{ab}:={\cal D}_a
          t_b$, and for which we require that $\chi_{ab}=\chi_{(ab)}$. 
\item{ii.} For vector fields $v^a$ on $\Sigma$ we require that $({\cal D}_a 
           v^e)P^b_e=D_av^b$, where $D_a$ is the Levi-Civita covariant 
           derivation on $T\Sigma$ determined by $h_{ab}$. \par
\ni
Then for any section $X^a=Nt^a+N^a$ of $V(\Sigma)$ we have ${\cal D}_eX^a=
(t^aD_eN+D_eN^a)+(\chi_e{}^at_b-\chi_{eb}t^a)X^b$. Thus it seems natural to
extend $D_e$ from the sections of $T\Sigma$ (i.e. of $\Theta(T\Sigma)$) to 
any section of $V(\Sigma)$ by requiring $D_et^a=0$, since then both ${\cal 
D}_e$ and $D_e$ would be defined on the same vector bundle and we could 
compare them. For the Levi-Civita derivation extended in this way we have 
$D_e P^a_b=0$, $D_eg_{ab}=0$ and

$$
{\cal D}_eX^a=D_eX^a+\Bigl(\chi_e{}^at_b-\chi_{eb}t^a\Bigr)X^b.  \eqno(3.2.1)
$$
\ni
Thus, for given $\Theta$, the covariant derivation ${\cal D}_e$ is completely 
determined by $g_{ab}$ and $\chi_{ab}$; i.e. for given $\iota$, ${\cal D}_e$ 
is completely determined by the initial data set. Suppose for a moment that 
$\Sigma$ is a spacelike hypersurface in a Lorentzian spacetime $(M,g_{ab})$, 
$\nabla_e$ is the four dimensional Levi-Civita covariant derivation and 
define ${\cal D}_a:=P^b_a\nabla_b$, the so-called 3-dimensional Sen operator 
[6]. Obviously ${\cal D}_a$ is well defined on any tensor field defined on 
the submanifold $\Sigma$, it annihilates the spacetime metric and satisfies 
the requirements i. and ii. above. It is easy to prove the converse of this 
statement, namely that the differential operator on the restriction to 
$\Sigma$ of the spacetime tangent bundle satisfying i. and ii. and 
annihilating the spacetime metric is unique. Thus we call the connection 
satisfying i. and ii. the real Sen connection on $V(\Sigma)$. The 
contraction of (3.2.1) with $t_a$ and the projection of it to $\Theta(T
\Sigma)$, respectively, are

$$\eqalignno{
\bigl({\cal D}_eX^a\bigr)t_a&=D_eN-\chi_{ea}N^a, &(3.2.2)\cr
\bigl({\cal D}_eX^a\bigr)P^b_a&=D_eN^b+N\chi^b{}_e.  &(3.2.3)\cr}
$$
\ni
Thus ${\cal D}_e$ can also be considered as a covariant derivation on the 
bundle of the pairs $(N,N^a)$ on $\Sigma$, the Whitney sum of the trivial 
real line bundle $\Sigma\times{\bf R}$ and $T\Sigma$.\par 
         Next calculate the action of the commutator of two ${\cal D}_e$'s 
on functions and on sections of $V(\Sigma)$:

$$\eqalignno{
\Bigl({\cal D}_e{\cal D}_f-{\cal D}_f{\cal D}_e\Bigr)\phi=&-2\chi^b{}_{[e} 
  t_{f]}{\cal D}_b\phi, &(3.2.4)\cr 
\Bigl({\cal D}_e{\cal D}_f-{\cal D}_f{\cal D}_e\Bigr)X^a=&-2\chi^b{}_{[e} 
  t_{f]}{\cal D}_bX^a- \Bigl(R^a{}_{bef}+\chi^a{}_e\chi_{bf}-\chi^a{}_f\chi 
  _{be}\Bigr)X^b-\cr
&-\Bigl(t^a\bigl(D_e\chi_{fb}-D_f\chi_{eb}\bigr)-t_b\bigl(D_e\chi^a{}_f-D_f
  \chi^a{}_e\bigr)\Bigr)X^b, &(3.2.5)\cr}
$$
\ni
where $R^a{}_{bef}$ is the curvature tensor of the Levi-Civita connection 
of $(\Sigma,h_{ab})$. Then one can read off the curvature and the `torsion' 
of the Sen connection: 

$$\eqalignno{
F^a{}_{bef}:=&R^a{}_{bef}+\chi^a{}_e\chi_{bf}-\chi^a{}_f\chi_{be}+\cr
+&t^a\bigl(D_e\chi_{fb}-D_f\chi_{eb}\bigr)-t_b\bigl(D_e\chi_f{}^a-D_f
\chi_e{}^a\bigr), &(3.2.6)\cr
T^e{}_{ab}:=&2\chi^e{}_{[a}t_{b]}.  &(3.2.7)\cr}
$$
\ni
Thus $F^a{}_{b\alpha\beta}$ represents the Gauss and Codazzi tensors, built 
from the initial data $h_{ab}$ and $\chi_{ab}$, appearing in the 3+1
decomposition of the curvature tensor of a Lorentzian spacetime. Namely, if 
$\Sigma$ is a spacelike hypersurface in $(M,g_{ab})$ and ${}^MR^a{}_{bcd}$ 
is the spacetime curvature tensor then $F^a{}_{bef}={}^MR^a{}_{bcd}P^c_e
P^d_f$. Note that $F^a{}_{b\alpha\beta}$ is the curvature in the strict 
sense of differential geometry [3]; i.e. it is a globally defined $so(1,3)$ 
Lie algebra valued 2-form on $\Sigma$. On the other hand, $T^e{}_{ab}$ is 
{\it not} a torsion in the strict sense, because the torsion is defined 
only for connections on principal bundles that are reduced subbundles of 
the linear frame bundle of the base manifold; i.e. if there is a soldering 
form. The true torsion, the pull back to the base manifold of the covariant 
exterior derivative of the soldering form, is always a vector valued 2-form 
on the base manifold. Here $T^e{}_{ab}$ is {\it not} such a 2-form {\it on} 
$\Sigma$, its projection to $\Sigma$ is zero.\par
      If $e^a_{\ua}$, $\zeta^{\ua}_a$ is a pair of dual $g_{ab}$-orthonormal 
frame fields then we can define the connection coefficients of the Sen 
connection with respect to these frames by $\Gamma^{\ua}_{\alpha{\ub}}:=
\zeta^{\ua}_e{\cal D}_\alpha e^e_{\ua}$. These form a globally defined 
$so(1,3)$ matrix Lie algebra valued 1-form on $\Sigma$, and the tetrad 
components of the curvature in its `internal indices', $F^{\ua}{}_{{\ub}
\alpha\beta}:=\zeta^{\ua}_ae^b_{\ub}F^a{}_{b\alpha\beta}$, are built up 
from the connection components $\Gamma^{\ua}_{\alpha{\ub}}$ in the well 
known manner.  \par
        Finally, let us consider the behaviour of the various quantities 
under conformal rescalings. For any function $\Omega:\Sigma\rightarrow(0,
\infty)$ the conformal rescaling of the fibre metric, $g_{ab}\mapsto\hat 
g_{ab}:= \Omega^2g_{ab}$, determines the rescaling of the spatial metric: 
$h_{ab} \mapsto\hat h_{ab}:=\Omega^2h_{ab}$, but it doesn't determine the 
rescaling of $\chi_{ab}$. However, recalling how the extrinsic curvature of 
a spacetime hypersurface behaves under a conformal rescaling of the {\it 
spacetime} metric, the new $\chi_{ab}$ is expected to depend on an 
additional independent function $\dot\Omega:\Sigma\rightarrow{\bf R}$ too, 
and we {\it define} the new $\chi_{ab}$ by $\hat\chi_{ab}:=\Omega\chi_{ab}+
\dot\Omega h_{ab}$. If, for the sake of later convenience, we define 
$\Upsilon_e:=D_e(\ln\Omega)$ and $\omega:=\Omega^{-1}\dot\Omega$, then the 
behaviour of the Levi-Civita and Sen derivations, respectively, are

$$\eqalignno{
\hat D_eX^a&=D_eX^a+\Bigl(P^a_e\Upsilon_b+P^a_b\Upsilon_e-h_{eb}h^{af}
  \Upsilon_f\Bigr)X^b, &(3.2.8)\cr
\hat{\cal D}_eX^a&={\cal D}_eX^a+\Bigl(P^a_e\Upsilon_b+P^a_b\Upsilon_e-
  h_{eb}h^{af}\Upsilon_f\Bigr)X^b+\omega\bigl(P^a_et_b-t^ah_{eb}\bigr)X^b.
  &(3.2.9)\cr}
$$
\ni
One can now calculate the conformal behaviour of the curvature of the 
Levi-Civita connection, of the `torsion' and of the curvature of the Sen 
connection:

$$\eqalignno{
\Omega^2\hat R^{ab}{}_{cd}&=R^{ab}{}_{cd}+4P^{[a}_{[c}\Bigl(D_{d]}\Upsilon
 ^{b]}-\Upsilon_{d]}\Upsilon^{b]}\Bigr)+P^{ab}_{cd}\Upsilon_e\Upsilon^e,
 &(3.2.10)\cr
\hat T^e{}_{ab}&=T^e{}_{ab}+2\omega P^e_{[a}t_{b]}, &(3.2.11)\cr
\Omega^2\hat F^{ab}{}_{cd}&=F^{ab}{}_{cd}+4P^{[a}_{[c}\Bigl(\bigl(D_{d]}
 \Upsilon^{b]}-\Upsilon_{d]}\Upsilon^{b]}+\chi^{b]}{}_{d]}\omega\bigr)+
 t^{b]}\bigl(D_{d]}\omega-\Upsilon_{d]}-\chi_{d]}{}^e\Upsilon_e\bigr) 
 \Bigr)+\cr
&+P^{ab}_{cd}\bigl(\Upsilon_e\Upsilon^e+\omega^2\bigr), &(3.2.12)\cr}
$$
\ni
where $P^{ab}_{cd}:=P^a_cP^b_d-P^a_dP^b_c$. If $e^a_{\ua}$, $\zeta^{\ua}_a$ 
is a pair of dual orthonormal bases, then, under the conformal rescaling, 
they must be rescaled as $e^a_{\ua}\mapsto\hat e^a_{\ua}:=\Omega^{-1}e^a 
_{\ua}$, $\zeta^{\ua}_a\mapsto\hat\zeta^{\ua}_a:=\Omega\zeta^{\ua}_a$. 
Thus the behaviour of the connection coefficients and the curvature 
components in such a basis are

$$\eqalignno{
\hat\Gamma^{\ua}_{e{\ub}}&=\Gamma^{\ua}_{e{\ub}}+\zeta^{\ua}_a\Bigl(P^a_e
 \Upsilon_f-h_{ef}h^{ac}\Upsilon_c\Bigr)e^f_{\ub}+\omega\zeta^{\ua}_a
 \Bigl(P^a_et_f-t^ah_{ef}\Bigr)e^f_{\ub}, &(3.2.13)\cr \hat
F^{\ua}{}_{{\ub}cd}&=\zeta^{\ua}_ae^b_{\ub}\hat F^a{}_{bcd}, &(3.2.14) \cr}
$$
\ni
where $\hat F^a{}_{bcd}$ is given by (3.2.12).
\bigskip
\bigskip

\ni
{\bf 3.3 The Sen--Chern--Simons functional on $V(\Sigma)$}\par
\bigskip
\ni
Following the general prescription of subsection 2.1, we can introduce the 
Chern--Simons functional $Y[\Gamma]$, built from the real Sen connection 
on the trivializable vector bundle $V(\Sigma)$. We call $Y[\Gamma]$ the 
Sen--Chern--Simons functional. Using formulae (3.2.12-14) it is a lengthy 
but straightforward calculation to derive how $Y[\Gamma]$ transforms under 
conformal rescalings: 

$$
Y[\Gamma]-Y[\hat\Gamma]=\int_\Sigma D_a\Bigl(\varepsilon^{abc}\bigl(
\Upsilon_e+\omega t_e\bigr)e^e_{\ua}{\cal D}_b\zeta^{\ua}_c\Bigr){\rm d}
\Sigma, \eqno(3.3.1)
$$
\ni
where ${\rm d}\Sigma:={1\over3!}\varepsilon_{\alpha\beta\gamma}$, the metric 
volume element determined by the 3-metric $h_{ab}$. Thus for compact $\Sigma$ 
the Sen--Chern--Simons functional is invariant with respect to rescalings that 
correspond to {\it spacetime} conformal rescalings; i.e. $Y[\Gamma]$ modulo 
$16\pi^2$ is a conformal invariant of the initial data set. Since by 
Lemma 3.1 there is a one-to-one correspondence between the homotopy classes 
of the global rotations $\Sigma\rightarrow SO(3)$ and the global Lorentz 
transformations $\Sigma\rightarrow SO_0(1,3)$, the integer $N$ in (2.2.1) 
can still be interpreted as twice the winding number of the global Lorentz 
transformation. \par
         Since for fixed $\iota$ the real Sen connection is completely 
determined by $h_{ab}$ and $\chi_{ab}$, $Y[\Gamma]$ can also be considered 
as a {\it second} order functional of the frame field $e^a_{\ua}$ and a 
first order functional of $\chi_{ab}$. Similarly to the Riemannian case, 
any variation $\delta h_{ab}$ of the 3-metric yields a variation $\delta_1
\Gamma^{\ua}_{a{\ub}}$ of the connection coefficients and an unspecified 
{\it small} gauge transformation, and any variation $\delta\chi_{ab}$ 
yields a variation $\delta_2\Gamma^{\ua}_{a{\ub}}$. Thus the variational 
derivatives of $Y[\Gamma]$ with respect to $h_{ab}$ and $\chi_{ab}$ are 
well defined, and, using the general formula (2.1.4), these derivatives 
can be calculated. Since by Lemma 3.1 the pure boost gauge transformations 
are all small, these calculations can be carried out in the time gauge, 
where the formulae are considerably simpler. The results are 

$$\eqalignno{
{\delta Y\over\delta\chi_{ab}}&=-8\sqrt{\vert h\vert}\varepsilon^{cd(a}D_c
   \chi^{b)}{}_d=\cr
&=:8\sqrt{\vert h\vert}H^{ab}, &(3.3.2)\cr
{\delta Y\over\delta h_{ab}}&=-4\sqrt{\vert h\vert}\Bigl\{Y^{ab}-\varepsilon
   ^{cd(a}\Bigl(D_c\bigl(\chi\chi^{b)}{}_d-\chi^{b)e}\chi_{ed}\bigr)-
   {1\over2}\chi^{b)}{}_c\bigl(D_e\chi^e{}_d-D_d\chi\bigr)\Bigr)+H^{e(a}
   \chi^{b)}{}_e\Bigr\}=\cr
&=:-4\sqrt{\vert h\vert}\Bigl(B^{ab}+H^{e(a}\chi^{b)}{}_e\Bigr). &(3.3.3) \cr}
$$
\ni
Here $Y_{ab}:=-\varepsilon_{cd(a}D^cR^d{}_{b)}$, the Cotton--York tensor of 
the intrinsic 3-geometry; and $H_{ab}$ would play the role of the magnetic 
part of the Weyl curvature of the spacetime $(M,g_{ab})$ if $\Sigma$ were a 
spacelike hypersurface in $M$. Both $H_{ab}$ and $B_{ab}$ are symmetric and 
trace free. Although, by (3.3.1), $Y[\Gamma]$ is invariant with respect to 
any {\it finite} conformal rescaling, by (3.3.2) and (3.3.3) it is easy to 
prove directly its invariance with respect to {\it infinitesimal} conformal 
rescalings: If $(\Omega(t),\dot\Omega(t))$ is a 1-parameter family of 
conformal factors such that $\Omega(0)=1$ and $\dot\Omega(0)=0$, then 

$$
\delta Y[\Gamma]:=\Bigl({{\rm d}\over{\rm d}t}Y[\Gamma(t)]\Bigr)\vert_{t=0}
=\int_\Sigma\Bigl\{{\delta Y\over\delta h_{ab}}2\delta\Omega h_{ab}+{\delta 
Y\over\delta\chi_{ab}}\bigl(\delta\Omega\chi_{ab}+\delta\dot\Omega h_{ab}
\bigr)\Bigr\}{\rm d}\Sigma=0,\eqno(3.3.4)
$$
\ni
where $\delta\Omega:=({{\rm d}\over{\rm d}t}\Omega(t))_{t=0}$ and $\delta\dot
\Omega:=({{\rm d}\over{\rm d}t}\dot\Omega(t))_{t=0}$. We give a geometric 
characterization of the stationary points of the Sen--Chern--Simons 
functional, $B_{ab}=0$ and $H_{ab}=0$, in section four. \par
\bigskip
\bigskip

\ni
{\bf 3.4 The Ashtekar--Chern--Simons functional on ${}^\pm\Lambda^2(\Sigma)$}
\par
\bigskip
\ni
Next we are constructing another representation of the gauge group, 
$SO_0(1,3)$, and the associated vector bundle. This will be the 
self-dual/anti-self-dual representation. We will see that the Chern--Simons 
functional constructed in this vector bundle is {\it not} invariant with 
respect to the conformal behaviour introduced in the second subsection. 
Thus the conformal invariance depends on the actual representation too. \par
     To start with, let $\Lambda^2(\Sigma)$ be the vector bundle of 2-forms 
on the fibres of $V(\Sigma)$; i.e. the fibre of $\Lambda^2(\Sigma)$ over a 
point $p\in\Sigma$ is $V^*_p\wedge V^*_p$. $\Lambda^2(\Sigma)$ is a 
trivializable, real vector bundle over $\Sigma$. The fibre metric $g_{ab}$ 
on $V(\Sigma)$ defines a fibre metric on $\Lambda^2(\Sigma)$ by $\langle
\alpha,\beta\rangle:=2g^{ac}g^{bd}\alpha_{ab}\beta_{cd}$, for any $\alpha
_{ab}=\alpha_{[ab]}$ and $\beta_{ab}=\beta_{[ab]}$. If $\zeta^{\ua}_a$, 
${\ua}=0,...,3$, is a basis in $V^*_p$ (or a global frame field for $V^*(
\Sigma)$), then $\zeta^{\ua}_{[a}\zeta^{\ub}_{b]}$, ${\ua}<{\ub}$, form a 
basis for $V^*_p$ (or in $\Lambda^2(\Sigma)$), and $\langle \zeta^{\ua}
\wedge\zeta^{\ub},\zeta^{\uc}\wedge\zeta^{\ud}\rangle=g^{{\ua}{\uc}}g^{{\ub}
{\ud}}-g^{{\ua}{\ud}}g^{{\ub}{\uc}}$. Thus if $\zeta^{\ua}_a$ is 
$g_{ab}$-orthonormal, then $\{\zeta^0\wedge\zeta^{\bi}, \zeta^{\bj}\wedge
\zeta^{\bk}\}$, ${\bi},{\bj},{\bk},...=1,2,3$, is $\langle,
\rangle$-orthonormal and $\langle\zeta^0\wedge\zeta^{\bi},\zeta^0\wedge
\zeta^{\bi}\rangle=-1$ and $\langle\zeta^{\bi}\wedge\zeta^{\bj},\zeta^{\bi}
\wedge\zeta^{\bj}\rangle=1$; i.e. the signature of $\langle,\rangle$ is 
$(- - - + + +)$. \par
          Let $\varepsilon_{abcd}$ be the $g_{ab}$-volume form on the fibres 
of $V(\Sigma)$, and introduce the duality mapping in the standard way: 
$\ast:\Lambda^2(\Sigma)\rightarrow\Lambda^2(\Sigma):$ $W_{ab}\mapsto\ast W
_{ab}:={1\over2}\varepsilon_{ab}{}^{cd}W_{cd}$. Then $\langle\ast\alpha,\beta
\rangle=\langle\alpha,\ast\beta\rangle$ and $\ast\ast=-{\rm Id}_{\Lambda^2(
\Sigma)}$. Thus the eigenvalues of the linear mapping $\ast$ are $\pm{\rm i}$, 
and hence its eigenvectors belong to $\Lambda^2(\Sigma)\otimes{\bf C}$, the 
complexification of $\Lambda^2(\Sigma)$. ${}^\pm W_{ab}:={1\over2}(W_{ab}\mp
{\rm i}\ast W_{ab})$ are called the self-dual/anti-self-dual part of the 
(real) 2-form $W_{ab}$. Thus the complexification of $\Lambda^2(\Sigma)$ can 
be decomposed in a natural way as the Withey sum of two of its subbundles: 
$\Lambda^2(\Sigma)\otimes{\bf C}={}^+\Lambda^2(\Sigma)\oplus{}^-\Lambda^2
(\Sigma)$. ${}^\pm\Lambda^2(\Sigma)$ are the bundle of 
self-dual/anti-self-dual 2-forms, respectively, over $\Sigma$. They are 
trivializable complex vector bundles of rank 3 over $\Sigma$. \par
           If $\zeta^{\ua}_a$ is any orthonormal dual global frame field 
then $\langle\ast\zeta^{\ua}\wedge\zeta^{\ub},\zeta^{\uc}\wedge\zeta^{\ud}
\rangle=-\epsilon^{\ua\ub\uc\ud}$, where $\epsilon^{\ua\ub\uc\ud}$ is the 
anti-symmetric Levi--Civita symbol, by means of which it is easy to 
calculate the self-dual/anti-self-dual part of the basis 2-forms. One has 
${}^\pm(\zeta^{\bi}\wedge\zeta^{\bj})=\pm{\rm i}\varepsilon^{\bi\bj}{}_{0
{\bk}}{}^\pm(\zeta^0\wedge\zeta^{\bk})$. Thus ${}^\pm\zeta^{\bi}_{ab}:=4
{}^\pm(\zeta^0_{[a}\zeta^{\bi}_{b]})$, ${\bi}=1,2,3$, form a basis in 
${}^\pm\Lambda^2(\Sigma)$ and $\langle{}^+\zeta^{\bi},{}^+\zeta^{\bj}
\rangle=8\eta^{\bi\bj}$, $\langle{}^+\zeta^{\bi},{}^-\zeta^{\bj}\rangle=0$. 
Therefore the self-dual and the anti-self-dual 2-forms are orthogonal to 
each other and, by $\overline{{}^+\zeta^{\bi}_{ab}}={}^-\zeta^{\bi}_{ab}$, 
they are also complex conjugate of each others. In the time gauge, i.e. if 
the pair of orthonormal global dual frame fields is $\{t^a,E^a_{\bi}\}$, 
$\{t_a,\vartheta^{\bi}_a\}$, the contraction of the normal section of 
$V(\Sigma)$ and the basis vectors of ${}^\pm\Lambda^2(\Sigma)$ is $t^a{}^\pm
\zeta^{\bi}_{ab}=\vartheta^{\bi}_b$. Therefore, {\it in the time gauge}, 
${}^\pm\Lambda^2(\Sigma)$ can be identified with the complexified tangent 
bundle $T\Sigma\otimes{\bf C}$ and its complex conjugate bundle, 
respectively, and $\vartheta^{\bi}_a$ can be chosen as a basis both in 
${}^+\Lambda^2(\Sigma)$ and in ${}^-\Lambda^2(\Sigma)$. \par
        The real Sen connection on $V(\Sigma)$ defines a unique connection on 
the vector bundles ${}^\pm\Lambda^2(\Sigma)$ by 

$$
{\cal D}_e{}^\pm W_{ab}:={1\over2}\Bigl({\cal D}_eW_{ab}\mp{{\rm i}\over2}
\varepsilon_{ab}{}^{cd}\bigl({\cal D}_eW_{cd}\bigr)\Bigr). \eqno(3.4.1)
$$
\ni
Thus if $\{e^a_{\ua}\}$, $\{\zeta^{\ua}_a\}$ is a pair of dual 
$g_{ab}$-orthonormal global frame fields in $V(\Sigma)$ and the 
corresponding connection coefficients of the real Sen connection are 
$\Gamma^{\ua}_{e{\ub}}:=\zeta^{\ua}_a{\cal D}_ee^a_{\ub}$, then the ${\cal 
D}_e$-derivative of the basis fields are ${\cal D}_e{}^\pm\zeta^{\bi}_{ab}=
-(\Gamma^{\bi}_{e{\bj}}\pm{\rm i}\Gamma^0_{e{\bk}}\varepsilon^{\bk\bi}
{}_{0{\bj}}){}^\pm\zeta^{\bj}_{ab}$; i.e. the connection coefficients of 
the connection (3.4.1) in the basis ${}^\pm\zeta^{\bi}_{ab}$ are 

$$
{}^\pm A^{\bi}_{e{\bj}}:=\Gamma^{\bi}_{e{\bj}}\pm{\rm i}\Gamma^0_{e{\bk}}
\varepsilon^{\bk\bi}{}_{\bj}, \eqno(3.4.2)
$$
\ni
where $\varepsilon_{\bi\bj\bk}:=\varepsilon_{0\bi\bj\bk}$. In the time gauge, 
when $\Gamma^{\bi}_{e{\bj}}$ reduces to the Ricci rotation coefficients 
$\gamma^{\bi}_{e{\bj}}$ of the spatial metric $h_{ab}$ in the spatial basis 
$\{E^a_{\bi}\}$ and $\Gamma^0_{e{\bk}}=-\chi_{ef}E^f_{\bk}$, ${}^\pm A^{\bi}
_{e{\bj}}$ become Ashtekar's connection coefficients [7]:

$$
{}^\pm A^{\bi}_{e{\bj}}=\gamma^{\bi}_{e{\bj}}\mp{\rm i}\chi_{ef}E^f_{\bk}
\varepsilon^{\bk\bi}{}_{\bj}. \eqno(3.4.3)
$$
\par
           Next let us consider the Chern--Simons functional built up from 
the connection ${}^\pm A^{\bi}_{e{\bj}}$ given by (3.4.2). $Y[{}^\pm A]$ 
can also be considered as a second order functional of $e^a_{\ua}$ and a 
first order functional of $\chi_{ab}$. Before calculating their variational 
derivatives, it seems useful to introduce the following notation:

$$\eqalignno{
V_{abcd}&:=\chi_{ac}\chi_{bd}-\chi_{ad}\chi_{bc}, \hskip 20pt
  V_{ab}:=V^e{}_{aeb}=\chi\chi_{ab}-\chi_{ae}\chi^e{}_b, \hskip 20pt
  V:=V^e{}_e=\chi^2-\chi_{ab}\chi^{ab}, &(3.4.4)\cr
J_a&:=D_b\chi^b{}_a-D_a\chi. &(3.4.5)\cr}
$$
\ni
The algebraic symmetries of $V_{abcd}$ and $V_{ab}$ are the same those of the 
Riemann and Ricci tensors, respectively. Then the tensors $B_{ab}$ and 
$H_{ab}$ of the previous subsection take the form:

$$\eqalignno{
B_{ab}&=-\varepsilon_{cd(a}D^c\Bigl(R^d{}_{b)}+V^d{}_{b)}\Bigr)+{1\over2}
   \chi^c{}_{(a}\varepsilon_{b)cd}J^d, &(3.4.6) \cr
H_{ab}&=-\varepsilon_{cd(a}D^c\chi^d{}_{b)}. &(3.4.7)\cr}
$$
\ni
Then the variational derivatives of $Y[{}^\pm A]$ with respect to $h_{ab}$ 
and $\chi_{ab}$, calculated most easily in the time gauge, are 

$$\eqalignno{
{\delta Y[{}^\pm A]\over\delta\chi_{ab}}=&2\sqrt{\vert h\vert}\Bigl(H^{ab}
   \mp{\rm i}\bigl(R^{ab}-{1\over2}Rh^{ab}+V^{ab}-{1\over2}Vh^{ab}\bigr)
   \Bigr), &(3.4.8)\cr
{\delta Y[{}^\pm A]\over\delta h_{ab}}=&-\sqrt{\vert h\vert}\Bigl(B^{ab}+
   \chi^{(a}{}_eH^{b)e}\Bigr)\mp \cr
&\mp{\rm i}\sqrt{\vert h\vert}\Bigl(\varepsilon^{ce(a}\varepsilon^{b)df}
    D_cD_d\chi_{ef}+\chi^{(a}{}_e\bigl(R^{b)e}-{1\over2}h^{b)e}R+V^{b)e}-
    {1\over2}h^{b)e}V\bigr)\Bigr). &(3.4.9)\cr}
$$
\ni
Using these formulae the variation of the Ashtekar--Chern--Simons functional 
under the infinitesimal conformal rescaling of the previous subsection can 
be given easily: 

$$
\delta Y[{}^\pm A]=\pm{{\rm i}\over2}\int_\Sigma\Bigl\{\delta\dot\Omega\bigl(
R+V\bigr) +4\delta\Omega\Bigl(D_aD^a\chi-D_aD_b\chi^{ab}-\chi_{ab}\bigl(
R^{ab}-{1\over2}Rh^{ab}+V^{ab}-{1\over2}Vh^{ab}\bigr)\Bigr)\Bigr\}{\rm d}
\Sigma. \eqno(3.4.10)
$$
\ni
Thus $Y[{}^\pm A]$ is {\it not} invariant even with respect to infinitesimal 
conformal rescalings. Thus the invariance of the functional depends not only 
on the connection on the principle bundle, but the actual representation 
$\rho$ of the structure group; i.e. the associated vector bundle too. \par
      The first term of the imaginary part on the right hand side of (3.4.9) 
can also be rewritten as 

$$
\varepsilon^{ce(a}\varepsilon^{b)df}D_cD_d\chi_{ef}=-2\varepsilon^{cd(a}
D_cH^{b)}{}_d-D_c\bigl(\varepsilon^{c(a}{}_eH^{b)e}\bigr)+{1\over2}h^{ab}
D_eJ^e-{1\over2}D^{(a}J^{b)}. \eqno(3.4.11)
$$
\ni
Thus for the stationary points of $Y[{}^\pm A]$ we have 

$$\eqalignno{
& H_{ab}=0, &(a.)\cr
&R_{ab}-{1\over2}Rh_{ab}+V_{ab}-{1\over2}Vh_{ab}=0, &(b.)\cr
&\varepsilon_{cd(a}D^c\bigl(R^d{}_{b)}+V^d{}_{b)}\bigr)={1\over2}\chi^c
  {}_{(a}\varepsilon_{b)cd}J^d, &(c.)\cr
&D_{(a}J_{b)}=h_{ab}D_eJ^e. &(d.)\cr}
$$
\ni
Now b. implies $R_{ab}+V_{ab}=0$ and d. implies that $D_{(a}J_{b)}=0$.
We will show that these two, together with $H_{ab}=0$, imply the vanishing 
of $J_a$. ($B^{ab}=0$, i.e. c., will not be used in what follows.) First we 
show that $J_a$ is constant. By $H_{ab}=0$ we have $D_{[a}\chi_{b]c}={1
\over2}h_{c[a}J_{b]}$, and, using $R_{ab}+V_{ab}=0$, a straightforward 
calculation shows that $D_{[a}J_{b]}={1\over2}D_{[a}J_{b]}$, i.e. $J_a$ is, 
in fact, constant. Then taking the divergence of b., we get $\chi^b{}_{[a}
J_{b]}=0$. Taking the divergence again and using $D_aJ_b=0$ we finally get 
$J_aJ^a=0$, i.e. by the definiteness of $h_{ab}$, that $J_a=0$. But $R_{ab}
+V_{ab}=0$ and $D_{[a}\chi_{b]c}=0$ together is just the Gauss--Codazzi 
condition for the local isometric imbeddability of $(\Sigma,h_{ab},\chi
_{ab})$ in a {\it flat} spacetime with first and second fundamental forms 
$h_{ab}$ and $\chi_{ab}$, respectively. \par
\bigskip
\bigskip

\ni
{\lbf 4. The criterion of non-contortedness of the initial data sets}\par
\bigskip
\ni
Let $\Sigma$ be an $n$ dimensional manifold, $n\geq3$, $h_{ab}$ a 
pseudo-Riemannian metric with signature $(p,q)$, $p+q=n$, and $\chi_{ab}$ a 
symmetric tensor field on $\Sigma$. The triple $(\Sigma,h_{ab},\chi_{ab})$ 
will be said to be {\it locally imbeddable} into the $n+1$ dimensional 
pseudo-Riemannian manifold $(M,g_{ab})$ as a non-null hypersurface if each 
point $p$ of $\Sigma$ has an open neighbourhood $U$ and there is an 
imbedding $\phi:U\rightarrow M$ such that $h_{ab}=\phi^\ast g_{ab}$ and 
$\chi_{ab}=\phi^\ast K_{ab}$, where $K_{ab}$ is the extrinsic curvature of 
$\phi(\Sigma)$ in $M$: $K_{ab}:=P^e_aP^f_b\nabla_et_f$. Here $t_a$ is the 
unit normal of $\phi(\Sigma)$, $g^{ab}t_at_b=\pm1$ and $P^a_b:=\delta^a_b\mp
t^at_b$, the projection to $\Sigma$ (the $n$ dimensional, or hypersurface, 
Kronecker delta). The triple will be called {\it non-contorted} [8] if it 
is locally imbeddable as a non-null hypersurface into some {\it conformally 
flat} geometry $(M,g_{ab})$. As is well known [8], for $n=3$ $(\Sigma,
h_{ab},\chi_{ab})$ is non-contorted iff the hypersurface twistor equation is 
completely integrable, i.e. it admits four linearly independent solutions.
\par
          In the present section we give an equivalent characterization of 
the non-contortedness in any dimensions greater than two by the vanishing of 
three tensor fields. In three dimensions one of these vanishes identically, 
while the others are precisely $B_{ab}$ and $H_{ab}$. Thus the stationary 
points of our conformal invariant are precisely the non-contorted initial 
data sets. In addition to the characterization of these stationary points, 
$B_{ab}=0$ and $H_{ab}=0$ provide a new criterion for the complete 
integrability of the hypersurface twistor equation. The main result of this 
section is the following statement:
\medskip
\ni
{\bf Proposition 4.1} The initial data set $(\Sigma,h_{ab},\chi_{ab})$ is 
non-contorted if and only if the following tensor fields vanish: 

$$\eqalignno{
E^{ab}{}_{cd}&:=C^{ab}{}_{cd}\pm\Bigl(V^{ab}{}_{cd}-{4\over(n-2)}P^{[a}_{[c}
  V^{b]}{}_{d]}+{2\over(n-1)(n-2)}P^{[a}_cP^{b]}_dV\Bigr)=0, &(4.1.i)\cr
H^{ijk}_{ab}&:=P^{[i}_cP^j_dP^{k]}_{(a}D^c\chi^d{}_{b)}=0, &(4.1.ii)\cr
B_{ab}{}^d&:={1\over(n-2)}\Bigl(D_{[a}L^d_{b]}\mp D_{[a}\bigl(V^d_{b]}-
  {1\over2(n-1)}VP^d_{b]}\bigr)\pm{(n-2)\over(n-1)}\chi^d{}_{[a}\bigl(D^c
  \chi_{b]c}-D_{b]}\chi\bigr)\Bigr)=0. &(4.1.iii)\cr}
$$
\ni
Here $L_{ab}:=-(R_{ab}-{1\over2(n-1)}Rh_{ab})$, $C^{ab}{}_{cd}:=R^{ab}{}_{cd}
+{4\over(n-2)}P^{[a}_{[c}L^{b]}_{d]}$ is the Weyl tensor of the metric 
$h_{ab}$, and $V_{abcd}$ and $V_{ab}$ are defined by (3.4.4). The sign $\pm$ 
corresponds to the sign of the length of the normal of $\Sigma$ in the 
imbedding: $g_{ab}t^at^b=\pm1$. 
\medskip
\ni
{\it Proof:}
First suppose that $(\Sigma,h_{ab},\chi_{ab})$ is locally imbedded into the 
conformally flat $(M,g_{ab})$ and for the sake of simplicity we identify 
$\Sigma$ with its $\phi$-image in $M$. Let $\tilde g_{ab}$ be a flat metric 
on $M$ such that $g_{ab}=\Omega^2\tilde g_{ab}$ for some positive function 
$\Omega$ on $M$, and let $\tilde\nabla_a$ be the corresponding flat 
Levi-Civita covariant derivation. Since $(M,\tilde g_{ab})$ is flat, 
there exist $(n+1)$ linearly independent 1-form fields $K_a$ which are 
constant with respect to the flat connection: $\tilde\nabla_aK_b=0$. Let 
$\nabla_a$ be the covariant derivation associated with the conformally flat 
metric $g_{ab}$. If $C^a_{eb}X^b:=(\nabla_e-\tilde\nabla_e)X^a$ then
 
$$\eqalign{
C^a_{eb}&=2\delta^a_{(e}\tilde\nabla_{b)}\ln\Omega-\tilde g_{eb}\tilde
  g^{af}\tilde\nabla_f\ln\Omega=\cr
  &=2\delta^a_{(e}\nabla_{b)}\ln\Omega-g_{eb}g^{af}\nabla_f\ln\Omega,\cr}
\eqno(1)
$$
\ni
and the Riemann tensor ${}^MR^{ab}{}_{cd}$ of the connection $\nabla_e$ takes 
the form 
 
$$
\Omega^2 \, {}^MR^{ab}{}_{cd}=4\delta^{[a}_{[c}\tilde\nabla_{d]}\tilde\nabla
   ^{b]}\ln\Omega -4\delta^{[a}_{[c}\tilde\nabla_{d]}\ln\Omega\tilde
   \nabla^{b]}\ln\Omega +2\delta^a_{[c}\delta^b_{d]}\tilde\nabla_e\ln\Omega
   \tilde\nabla^e\ln\Omega. \eqno(2)
$$
\ni
Here the raising and lowering of indices on the right hand side is defined 
by the flat metric, while ${}^MR^{ab}{}_{cd}=g^{be}{}^MR^a{}_{ecd}$. In 
what follows we rewrite every quantity using only the conformally flat 
metric $g_{ab}$. In particular, in terms of $\nabla_e$, eq.(2) takes the 
form 

$$
{}^MR^{ab}{}_{cd}= 4\delta^{[a}_{[c}\nabla_{d]}\nabla^{b]}\ln\Omega+
   4\delta^{[a}_{[c}\nabla_{d]}\ln\Omega\nabla^{d]}\ln\Omega-
   2\delta^a_{[c}\delta^b_{d]}\nabla_e\ln\Omega\nabla^e\ln\Omega,\eqno(3)
$$
\ni
and the $\tilde\nabla$-constant 1-form fields satisfy

$$
\nabla_aK_b=-2K_{(a}\nabla_{b)}\ln\Omega+g_{ab}K^e\nabla_e\ln\Omega.
\eqno(4)
$$
\ni
Let us define $\bar k_a:=P^e_aK_e$ and $\bar\xi:=t^aK_a$, by means of which
$K_a=\bar k_a\pm\bar\xi t_a$. From eq.(4) we have

$$\eqalignno{
D_a\bar k_b\pm\bar\xi\chi_{ab}&=-2\bar k_{(a}\Upsilon_{b)}+h_{ab}\Bigl(\bar
   k^e\Upsilon_e\pm\bar\xi\Omega^{-1}\dot\Omega\Bigr)&(5)\cr
D_a\bar\xi-\chi_{ab}\bar k^b&=-\bar k_a\Omega^{-1}\dot\Omega-\bar\xi
   \Upsilon_a. &(6)\cr}
$$
\ni
Here $D_a$ is the Levi-Civita covariant derivation on $\Sigma$, $\dot
\Omega:=t^e\nabla_e\Omega$ and $\Upsilon_a:=D_a\ln\Omega$. Then by (5) and 
(6) $D_a(\bar k_e\bar k^e\pm\bar\xi^2)=-2\Upsilon_a(\bar k_e\bar k^e\pm\bar
\xi^2)$, which implies that $\Omega^2(\bar k_e\bar k^e\pm\bar\xi^2)={\rm
const}$. Thus it seems natural to introduce the following notations:

$$
k_a:=\Omega\bar k_a,\hskip 12pt
\xi:=\Omega\bar\xi, \hskip 12pt
\omega:=\Omega^{-1}\dot\Omega. \eqno(7)
$$
\ni
Then by (5)-(7) and the definition of $\Upsilon_a$ we have

$$\eqalignno{
D_ak_b\pm\xi\chi_{ab}&=-k_a\Upsilon_b+h_{ab}\Bigl(k^e\Upsilon_e
                       \pm\omega\xi\Bigr) &(8)\cr
D_a\xi-\chi_{ab}k^b&=-\omega k_a, &(9)\cr
D_a\Upsilon_b&=D_b\Upsilon_a. &(10)\cr}
$$
\ni
Equations (8-10) form a system of partial differential equations for 
$k_a$ and $\xi$, whose conditions of integrability are

$$\eqalignno{
0=\Bigl(D_aD_b-D_bD_a\Bigr)\xi=&2\Bigl(D_{[a}\chi_{b]c}+h_{c[a}\bigl(D_{b]}
  \omega+\Upsilon_{b]}\omega-\chi_{b]e}\Upsilon^e\bigr)\Bigr)k^c, &(11)\cr
R^{cd}{}_{ab}k_d=-\Bigl(D_aD_b-D_bD_a\Bigr)k^c=&\pm2\Bigl(D_{[a}\chi_{b]}{}
  ^c+P^c_{[a}\bigl(D_{b]}\omega+\omega\Upsilon_{b]}-\chi_{b]e}\Upsilon^e
  \bigr)\Bigr)\xi+\cr
 &+2\Bigl(\mp\chi^c{}_{[a}\chi^d{}_{b]}+2P^{[c}_{[a}D_{b]}\Upsilon^{d]}+
  P^{[c}_{[a}\Upsilon_{b]}\Upsilon^{d]}\pm\cr
 &\pm\omega P^{[c}_{[a}\chi^{d]}{}_{b]}-{1\over2}P^c_{[a}P^d_{b]}
  \bigl(\Upsilon_e\Upsilon^e\pm\omega^2\bigr)\Bigr)k_d. &(12)\cr}
$$
\ni
Applying $P^a_b$ to eq. (3) we obtain 

$$\eqalignno{
{}^MR_{ijkl}P^i_aP^j_bP^k_cP^l_d=2&\Bigl(h_{a[c}D_{d]}\Upsilon_b-
   h_{b[c}D_{d]}\Upsilon_a+h_{a[c}\Upsilon_{d]}\Upsilon_b-h_{b[c}\Upsilon
   _{d]}\Upsilon_a\pm\cr
  &\pm\omega\bigl(h_{a[c}\chi_{d]b}-h_{b[c}\chi_{d]a}\bigr)-h_{a[c}h_{d]b}
   \bigl(\Upsilon_e\Upsilon^e\pm\omega^2\bigr)\Bigr), &(13)\cr
{}^MR_{ajkl}t^aP^j_bP^k_cP^l_d=-&2h_{b[c}\Bigl(D_{d]}\omega+\omega
   \Upsilon_{d]}-\chi_{d]e}\Upsilon^e\Bigr). &(14)\cr}
$$
\ni
On the other hand the $(n+1)$ dimensional curvature tensor can be
expressed in terms of the $n$ dimensional curvature tensor and the
extrinsic curvature, and hence we finally have
 
$$\eqalignno{
R^{ab}{}_{cd}\pm2\chi^a{}_{[c}\chi^b{}_{d]}&=4P^{[a}_{[c}D_{d]}\Upsilon^{b]}
  +4P^{[a}_{[c}\Upsilon_{d]}\Upsilon^{b]}\pm4\omega P^{[a}_{[c}\chi^{b]}{}
  _{d]}-2P^a_{[c}P^b_{d]}\bigl(\Upsilon_e\Upsilon^e\pm\omega^2\bigr),&(15)\cr
D_c\chi_{db}-D_d\chi_{cb}&=-2h_{b[c}\Bigl(D_{d]}\omega+\omega\Upsilon_{d]}-
  \chi_{d]e}\Upsilon^e\Bigr). &(16)\cr}
$$
\ni
Thus by (15), (16) the integrability conditions (11,12) of the system 
(8,9) are satisfied identically. Equations (15,16) contain two kinds of 
information: One is already in the form of conditions on $(h_{ab},\chi_{ab}
)$. The other kind is a system of partial differential equations on $(\omega,
\Upsilon_a)$, which we obtain by contraction eqs.(15,16), namely eqs. (18,19) 
below, and which is again overdetermined. By writing down the integrability 
conditions to this latter system, we will finally arrive at the complete 
characterization of non-contortedness.\par
       The contractions of (15,16) are
 
$$\eqalignno{
R_{bd}\pm\bigl(\chi\chi_{bd}-\chi_{bc}\chi^c{}_d\bigr)&=(n-2)
   \Bigl(D_b\Upsilon_d+\Upsilon_b\Upsilon_d\pm\omega\chi_{bd}-h_{ab}
   \Upsilon_e\Upsilon^e\Bigr)+\cr
 &+h_{bd}\Bigl(D_e\Upsilon^e\pm\Omega\chi\mp(n-1)\omega^2\Bigl), &(17)\cr
R\pm\Bigl(\chi^2-\chi_{ab}\chi^{ab}\Bigr)&=(n-1)\Bigl(2D_e\Upsilon^e-(n-2)
  \Upsilon_e\Upsilon^e\pm2\omega\chi\mp n\omega^2\Bigr) &(18)\cr
D_c\chi^c{}_d-D_d\chi&=-(n-1)\Bigl(D_d\omega+\omega\Upsilon_d-\chi_{de}
   \Upsilon^e\Bigr). &(19)\cr}
$$
\ni
Then by (17,18)
 
$$\eqalign{
L_{bd}&=\pm\Bigl(\bigl(\chi\chi_{bd}-\chi_{bc}\chi^c{}_d\bigr)-{1\over2(n-1)}
   h_{bd}\bigl(\chi^2-\chi_{ac}\chi^{ac}\bigr)\Bigr)-\cr
  &-(n-2)\Bigl(D_b\Upsilon_d+\Upsilon_b\Upsilon_d\pm\omega\chi_{bd}-
   {1\over2}h_{bd}\bigl(\Upsilon_e\Upsilon^e\pm\omega^2\bigr)\Bigr).\cr}
\eqno(20)
$$
\ni
Then substituting (20) back into eq.(15) and using the definition of the Weyl 
tensor we obtain
 
$$
E^{ab}{}_{cd}:=C^{ab}{}_{cd}\pm\Bigl(V^{ab}{}_{cd}-{4\over(n-2)}P^{[a}_{[c}
V^{b]}{}_{d]}+{2\over(n-1)(n-2)}P^a_{[c}P^b_{d]}V\Bigr)=0.\eqno(21)
$$
\ni
$E_{abcd}$ plays the role of the Weyl tensor for the initial data sets.
If $n=3$ then $C_{abcd}$ and the term involving $V^{ab}{}_{cd}$ in the 
expression for $E_{abcd}$ are separately zero identically. Also, in this 
case, equations (15) and (20) are equivalent. Next consider equation (16) 
and its contraction, eq. (19). By means of (19) eq.(16) can be rewritten as

$$
D^c\chi^d{}_b-D^d\chi^c{}_b={2\over(n-1)}P^{[c}_b\Bigl(D_e\chi^{d]e}-
D^{d]}\chi\Bigr).\eqno(22)
$$
\ni
Contracting this equation with $P^{ijk}_{cda}:=3!P^i_{[c}P^j_dP^k_{a]}$ we 
obtain
 
$$
{1\over(n-1)}P^{ijk}_{abd}\Bigl(D_c\chi^{cd}-D^d\chi\Bigr)-P^{ijk}_{cd[a}D^c
\chi^d{}_{b]}=P^{ijk}_{cd(a}D^c\chi^d{}_{b)}. \eqno(23)
$$
\ni
Since its left hand side is antisymmetric in $ab$ and its right hand side
is symmetric in $ab$, they must vanish separately:
 
$$\eqalignno{
A^{ijk}_{ab}&:=\Bigl({1\over(n-1)}P^{ef}_{cd}P^{ijk}_{abf}+{1\over2}P^{ef}
  _{ab}P^{ijk}_{cdf}\Bigr)D^{[c}\chi^{d]}{}_e=0, &(24)\cr
H^{ijk}_{ab}&:=P^{ijk}_{cd(a}D^c\chi^d{}_{b)}=0. &(25)\cr}
$$
\ni
The possible independent contractions of $H^{ijk}_{ab}$ are

$$\eqalignno{
h^{ab}H^{ijk}_{ab}&=0 &(26)\cr
H^{ejk}_{eb}&=(n-1)\Bigl(D^{[j}\chi^{k]}{}_b-{1\over(n-1)}P^{[j}_b\bigl(D_e
  \chi^{k]e}-D^{k]}\chi\bigr)\Bigr). &(27)\cr}
$$
\ni
Thus by (27) $H^{ijk}_{ab}=0$ is equivalent to (22), and hence implies 
$A^{ijk}_{ab}=0$. Thus eq.(16) is equivalent to eq.(19) together with 
eq.(25). \par
      Next let us consider the contracted equations (19) and (20):
 
$$\eqalignno{
(n-1)D_b\omega&=-\bigl(D_c\chi^c{}_b-D_b\chi\bigr)-(n-1)\Bigl(\omega
   \Upsilon_b-\chi_{bc}\Upsilon^c\Bigr), &(28)\cr
(n-2)D_b\Upsilon_d&=-L_{bd}\pm\Bigl(V_{bd}-{1\over2(n-1)}h_{bd}V\Bigr)-(n-2)
   \Bigl(\Upsilon_b\Upsilon_d\pm\omega\chi_{bd}-{1\over2}h_{bd}\bigl(
   \Upsilon_e\Upsilon^e\pm\omega^2\bigr)\Bigr). &(29)\cr}
$$
\ni
These equations can be considered as a system of partial differential 
equations for $\Upsilon_b$ and $\omega$. Their integrability conditions 
are

$$\eqalignno{
0=\bigl(D_aD_b-D_bD_a)\omega=&{1\over12}(n-1)h_{ai}h_{bj}\Bigl(H^{eij}_{ef}
   \Upsilon^f-{1\over(n-2)}D^fH^{eij}_{ef}\Bigr), &(30)\cr
R^{cd}{}_{ab}\Upsilon_c=\bigl(D_aD_b-D_bD_a\bigr)\Upsilon^d=
   &\Bigl(-{4\over(n-2)}P^{[c}_{[a}L^{d]}_{b]}\mp\bigl(V^{cd}{}_{ab}-{4\over
   (n-2)}P^{[c}_{[a}V^{d]}_{b]}+\cr
&+{2\over(n-1)(n-2)}VP^c_{[a}P^d_{b]}\bigr)\Bigr)\Upsilon_c\mp{2
   \over(n-1)}\omega h_{aj}h_{bk}H^{fjk}_{fe}h^{ed}-\cr
&-{2\over(n-2)}\Bigl(D_{[a}L^d_{b]}\mp D_{[a}\bigl(V^d_{b]}-{1\over2(n-1)}V
   P^d_{b]}\bigr)\pm\cr
 &\pm{(n-2)\over(n-1)}\chi^d{}_{[a}\bigl(D^c\chi_{b]c}-D_{b]}\chi\bigr)
\Bigr). &(31)\cr}
$$
\ni
Thus by (22) and (27) the first condition is satisfied, while, using 
(21,22,27), the second can be rewritten as 

$$\eqalign{
B_{ab}{}^d:&={1\over(n-2)}\Bigl(D_{[a}L^d_{b]}\mp D_{[a}\bigl(V_{b]}^d-{1
  \over2(n-1)}VP^d_{b]}\bigl)\pm{(n-2)\over(n-1)}\chi^d{}_{[a}\bigl(D^c\chi
  _{b]c}-D_{b]}\chi\bigr)\Bigr)=\cr
&=-{1\over2}E_{ab}{}^{cd}\Upsilon_c\mp{1\over(n-1)}\omega h_{aj}h_{bk}H^{fjk}
  _{fe}h^{ed}=0.\cr}\eqno(32)
$$
\ni
Obviously, $B_{abd}=B_{[ab]d}$ and $B_{[abd]}=0$. 
Thus, to summarize, if $(\Sigma,h_{ab},\chi_{ab})$ is non-contorted then 
$E^{ab}{}_{cd}=0$, $H^{ijk}_{ab}=0$ and $B_{ab}{}^d=0$. \par 
       Conversely, let the initial data set $(\Sigma,h_{ab},\chi_{ab})$ 
satisfy the conditions i.-iii. of the proposition. We show that this data 
set can be imbedded locally into a conformally flat geometry. First let us 
consider the system of partial differential equations (28), (29) for $\omega$ 
and $\Upsilon_b$. Its integrability conditions are the equations (30) and 
(31), which, by the conditions i.-iii., are satisfied independently of 
$\omega$ and $\Upsilon_a$. Thus by the Darboux theorem the system (28), (29) 
is completely integrable: for any 1-form $\Upsilon_a(p_0)$ at a given point 
$p_0\in\Sigma$ and real number $\omega(p_0)$ there is a uniquely determined 
solution of the system (28), (29) whose value at $p_0$ is just the pair 
$\Upsilon_a(p_0)$, $\omega(p_0)$. Then by i. and ii. the pair $(\Upsilon_a,
\omega)$ is also a solution of the system of equations (15), (16). 
Next, for a given pair $(\Upsilon_a,\omega)$, let us consider the system of 
partial differential equations (8), (9) for $k_a$ and $\xi$. Its 
integrability conditions are (11) and (12), which, by (15) and (16), are 
identically satisfied independently of $k_a$ and $\xi$. Thus the system 
(8), (9) is completely integrable, and it has n+1 linearly independent 
solutions $(k^{\ba}_a,\xi^{\ba})$, $\ba=0,1,...,n$, specified in the 
following way. Let $\{x^\alpha\}$, $\alpha=1,...,n$, be a local coordinate 
system around $p_0\in\Sigma$ in which $h_{\alpha\beta}(p_0)=\eta_{\alpha
\beta}:={\rm diag}(1,...,1,-1,...,-1)$. (The number of +1's is $p$ and the 
number of -1's is $q$.) Then the components of the solution 1-forms $k^{\ba}
_a$ in this coordinate system at $p_0$ and the value of the $\xi^{\ba}$'s at 
$p_0$ are chosen to satisfy $k^0_\alpha=0$, $\xi^0=1$ and $k^\beta_\alpha=
\delta^\beta_\alpha$, $\xi^\beta=0$. \par
      In a sufficiently small neighbourhood $U^{\prime\prime}$ of $p_0$ 
the 1-form $\Upsilon_a$ is not only closed (by (29)), but exact. Thus 
there exists a strictly positive smooth function $\Omega:U^{\prime\prime}
\rightarrow(0,\infty)$ such that $\Upsilon_a=D_a\ln\Omega$. Then let us 
define the following rescaling: $\bar k^{\ba}_a:=\Omega^{-1}k^{\ba}_a$, 
$\bar\xi^{\ba}:=\Omega^{-1}\xi^{\ba}$ and define $\dot\Omega:=\Omega\omega$. 
Then $\bar k^{\ba}_a$ and $\bar\xi^{\ba}$, defined only on $U^{\prime
\prime}$, satisfy 

$$\eqalignno{
D_a\bar k_b\pm\bar\xi\chi_{ab}&=-2\bar k_{(a}\Upsilon_{b)}+h_{ab}\Bigl(\bar
   k^e\Upsilon_e\pm\bar\xi\Omega^{-1}\dot\Omega\Bigr)&(33)\cr
D_a\bar\xi-\chi_{ab}\bar k^b&=-\bar k_a\Omega^{-1}\dot\Omega-\bar\xi
   \Upsilon_a. &(34)\cr}
$$
\ni
By (33) $\bar k^{\ba}_a$ are closed 1-forms on $U^{\prime\prime}$. Thus 
in a sufficiently small open neighbourhood $U^\prime\subset U^{\prime
\prime}$ of $p_0$ they are exact too, and hence there exist smooth 
functions $\phi^{\ba}:U^\prime\rightarrow{\bf R}$ such that $\bar k^{\ba}_a
=D_a\phi^{\ba}$. Because of the special choice of the $k^{\ba}_a$ at $p_0$ 
there is an open neighbourhood $U\subset U^\prime$ of $p_0$ on which the 
rank of the mapping $\phi:=\{\phi^{\ba}\}:U\rightarrow{\bf R}^{n+1}$ is 
$n$, i.e. $\phi$ is an imbedding of $U$ into the n+1 dimensional manifold 
${\bf R}^{n+1}$ with the natural Descartes coordinates $x^{\ba}$. At the 
points of $\phi(U)\subset {\bf R}^{n+1}$ let us define the functions 
$\tilde g^{{\ba}{\bb}}(\phi(p)):=\Omega^2(\pm\bar\xi^{\ba}(p)\bar\xi^{\bb}
(p)+\phi^{\ba}_{,\alpha}(p)\phi^{\bb}_{,\beta}(p)h^{\alpha\beta}(p))$ 
$\forall p\in U$. By (33) and (34) these are constant on $\phi(U)$: 
$D_\mu\tilde g^{{\ba}{\bb}}=0$, and, because of the special choice of the 
independent solution 1-forms and functions $(k^{\ba}_a,\xi^{\ba})$ at 
$p_0$, $\tilde g^{\ba\bb}(\phi(p_0))=\eta^{\ba\bb}:={\rm diag}(\pm1,1,...,
1,-1,...,-1)$. Then extend $\tilde g^{\ba\bb}$ to ${\bf R}^{n+1}$ in a 
constant way. Thus ${\bf R}^{n+1}$ together with $\tilde g_{\ba\bb}$, the 
inverse of $\tilde g^{\ba\bb}$, is a (flat) pseudo-Euclidean geometry. 
Since by $\phi^{\ba}_{,\alpha}\tilde g_{\ba\bb}\bar\xi^{\bb}=0$ the 1-form
$\bar\xi^{\ba}\tilde g_{\ba\bb}$ annihilates every vecor tangent to $\phi
(U)$, this 1-form is a normal of $\phi(U)$ in ${\bf R}^{n+1}$; and its norm 
with respect to $\tilde g_{\ba\bb}$ is $\bar\xi^{\ba}\bar\xi^{\bb}\tilde 
g_{\ba\bb}=(\bar\xi^{\ba}\tilde g_{\ba\bc})(\bar\xi^{\bb}\tilde g_{\bb\bd})
\tilde g^{\bc\bd}=\pm\Omega^2(\bar\xi^{\ba}\bar\xi^{\bb}\tilde g_{\ba\bb})
^2$, i.e. $\bar\xi^{\ba}\bar\xi^{\bb}\Omega^2\tilde g_{\ba\bb}=\pm1$. Let us 
extend the function $\Omega$ from $\phi(U)$ onto ${\bf R}^{n+1}$ to be 
positive everywhere and satisfying $\bar\xi^{\ba}\partial_{\ba}\Omega=\dot
\Omega$, where $\partial_{\ba}$ is the partial derivative with respect to 
$x^{\ba}$. Then $g_{\ba\bb}:=\Omega^2\tilde g_{\ba\bb}$ is a conformally 
flat metric on ${\bf R}^{n+1}$ with respect to which $\bar\xi^{\ba}$ is a 
unit normal of $\phi(U)$. The pull back to $U$ of this metric is $\phi^{\ba}
_{,\alpha}\phi^{\bb}_{,\beta}g_{\ba\bb}=\Omega^2\phi^{\ba}_{,\alpha}\tilde 
g_{\ba\bc}\tilde g^{\bc\bd}\tilde g_{\bd\bb}\phi^{\bb}_{,\beta}=(\phi^{\ba}
_{,\alpha}g_{\ba\bc}\phi^{\bc}_{,\mu})h^{\mu\nu}(\phi^{\bd}_{,\nu}g_{\bd\bb}
\phi^{\bb}_{,\beta})$, implying that $\phi^{\ba}_{,\alpha}\phi^{\bb}_{,\beta}
g_{\ba\bb}=h_{\alpha\beta}$. Finally, let us calculate the pull back to $U$ 
of the extrinsic curvature of $\phi(U)$. The Christoffel symbols of $({\bf 
R}^{n+1},g_{\ba\bb})$ in the coordinates $x^{\ba}$ are $\Gamma^{\ba}_{\bc\bb}
=2\delta^{\ba}_{({\bb}}\partial_{{\bc})}\ln\Omega-g_{\bb\bc}g^{\ba\bd}
\partial_{\bd}\ln\Omega$, thus the pullback to $U$ of the extrinsic 
curvature is $\phi^{\ba}_{,\alpha}\phi^{\bb}_{,\beta}\nabla_{\ba}(\bar
\xi^{\bc}g_{\bc\bb})=\chi_{\alpha\beta}-\Omega^{-1}\dot\Omega h_{\alpha
\beta}-\phi^{\ba}_{,\alpha}\phi^{\bb}_{,\beta}\Gamma^{\bc}_{\ba\bb}g_{\bc
\bd}\bar\xi^{\bd}=\chi_{\alpha\beta}$. \sq\par
\bigskip

\ni
Under a conformal rescaling of the initial data set the conditions i.-iii. 
of the proposition are expected to be invariant. To check this, we should 
calculate the behaviour of the tensor fields $E^{ab}{}_{cd}$, $H^{ijk}_{ab}$ 
and $B_{ab}{}^d$ under conformal rescalings. The results are:

$$\eqalignno{
\hat E^a{}_{bcd}&=E^a{}_{bcd}, &(4.2)\cr
\hat H^{ijk}_{ab}&=\Omega^{-3}H^{ijk}_{ab}, &(4.3)\cr
\hat B_{abc}&=B_{abc}+{1\over2}E_{abc}{}^dD_d\ln\Omega \mp{1\over(n-1)}
   \Omega^{-1}\dot\Omega h_{aj}h_{bk}H^{fjk}_{fc}. &(4.4)\cr}
$$
\ni
Thus the conditions i.-iii. are, in fact, conformally invariant. \par
       Next let us consider the physically important special case of $n=3$. 
As we mentioned in the proof above, in three dimensions $E_{abcd}=0$ 
identically. Furthermore $A^{ijk}_{ab}$ is also zero identically and ii. is 
equivalent to $H_{ab}:={1\over3!}(-)^q\varepsilon_{ijk}H^{ijk}_{ab}=(-)^q
\varepsilon_{cd(a}D^c\chi^d{}_{b)}=0$, the vanishing of the conformal 
magnetic curvature. (Here $q$ is the number of -1's in the pseudo-euclidean 
form of $h_{ab}$.) Finally, 

$$\eqalign{
\varepsilon^{cd}{}_aB_{cdb}&=Y_{ab}\mp\varepsilon_{cd(a}\Bigl(
   D^c\bigl(\chi\chi^d{}_{b)}-\chi_{b)e}\chi^{ed}\bigr)-{1\over2}\chi_{b)}
   {}^c\bigl(D_e\chi^{de}-D^d\chi\bigr)\Bigr)\mp\cr
 &\mp{1\over2}H^{fjk}_{fc}\chi^c{}_j\varepsilon_{kab},\cr}\eqno(4.5)
$$
\ni
and therefore ii., iii. are equivalent to

$$\eqalignno{
H_{ab}&:={1\over3!}(-)^q\varepsilon_{ijk}H^{ijk}_{ab}=(-)^q\varepsilon_{cd(a}
 D^c\chi^d{}_{b)}=0, &(ii'.)\cr
B_{ab}&:=Y_{ab}\mp\varepsilon_{cd(a}\Bigl(D^c\bigl(\chi\chi^d{}_{b)}-
   \chi_{b)e}\chi^{ed}\bigr)-{1\over2}\chi_{b)}{}^c\bigl(D_e\chi^{de}-
   D^d\chi\bigr)\Bigr)=0. &(iii'.)\cr}
$$
\ni
Both $H_{ab}$ and $B_{ab}$ are traceless and symmetric, for negative definite 
$h_{ab}$ they are the tensors $H_{ab}$ and $B_{ab}$ introduced in subsection 
3.3, and if $\chi_{ab}=0$ (i.e. the initial data set is `time symmetric') 
then $H_{ab}$ vanishes and $B_{ab}$ reduces to the Cotton--York tensor. Thus 
we have proven the following corollary:
\medskip
\ni
{\bf Corollary} The three dimensional initial data set is non-contorted if 
and only if $B_{ab}=0$ and $H_{ab}=0$. \par
\medskip
\ni
The conformal behaviour of the symmetric traceless tensors $B_{ab}$ and 
$H_{ab}$ are:

$$\eqalignno{
\hat B_{ab}&=\Omega^{-1}\Bigl(B_{ab}\mp (-)^q\Omega^{-1}\dot\Omega H_{ab}
                        \Bigr), &(4.6)\cr
\hat H_{ab}&=H_{ab}. &(4.7)\cr}
$$
\ni
Thus, as is well known, $H_{ab}$ is a conformal invariant of the initial 
data set; and for `internal' conformal rescalings (i.e. when $\dot\Omega=0$) 
$B_{ab}$ transforms covariantly, i.e. it has definite conformal weight, 
namely -1. \par

\bigskip
\bigskip
\ni
{\lbf Acknowledgements}\par
\bigskip
\ni
We are grateful to Helmuth Urbantke, Paul Tod, Korn\'el Szlach\'anyi, 
Helmut Friedrich and J\"org Frauendiener for discussions. One of us 
(L.B.Sz.) is indebted to the Erwin Sch\"odinger Institute, Vienna, 
where most of the present work was done for hospitality, and to the 
Albert Einstein Institut f\"ur Gravitationsphysik, Potsdam, where a part 
of this paper was completed. This work was partially supported by the 
Hungarian Scientific Research Fund grant OTKA T016246, the B\"uro f\"ur 
Austauschprogramme mit Mittel- und Osteuropa, Projekt 23\"o10. 

\bigskip
\bigskip

\ni
{\lbf References}\par
\bigskip
\item{[1]} S.S. Chern, J. Simons, {\it Characteristic forms and geometrical 
           invariants}, Ann.Math. {\bf 99} 48 (1974)
\item{   } S.S. Chern, {\it On a conformal invariant of three dimensional
           manifolds}, in Aspects of Mathematics and its Applications, Ed. 
           J.A. Barroso,Elsevier Science Publishers B.V. (1986)

\item{[2]} R. Penrose, W. Rindler, Spinors and spacetime, Vol. 1, 
           Cambridge University Press, Cambridge 1982

\item{[3]} S. Kobayashi, K. Nomizu, Foundations of Differential Geometry,
           Vol. 1 and Vol. 2, Interscience, New York, 1964 and 1968 

\item{[4]} J. Baez, J.P. Muniain, Gauge Fields, Knots and Gravity, World 
           Scientific, Singapore, 1994

\item{[5]} M. Spivak, A Comprehensive Introduction to Differential Geometry,
	 Vol. 1, Publish or Perish, Inc., Houston, Texas, 1979

\item{[6]} A. Sen, {\it On the existence of neutrino `zero-modes' in vacuum 
           spacetimes}, J.Math.Phys. {\bf 22} 1781 (1981)

\item{[7]} A. Ashtekar, Lectures on Non-perturbative Canonical Gravity, 
           World Scientific, Singapore, 1991

\item{[8]} C.R. Lebrun, {\it ${\cal H}$-space with a cosmological constant}, 
           Proc.Roy.Soc.London A {\bf 380} 171 (1982)
\item{   } K.P. Tod, {\it Three-surface twistors and conformal embedding}, 
           Gen.Rel.Grav. {\bf 16} 435 (1984) 

\end